\documentclass[10pt, a4paper]{article}
\pdfoutput=1
\usepackage{amssymb,amsmath,amsfonts}
\usepackage[dvipsnames]{xcolor}
\usepackage{graphicx}
\usepackage{longtable}
\usepackage{verbatim}
\usepackage{color}
\usepackage{mdframed}
\usepackage{soul}
\usepackage{mathrsfs}
\usepackage{amsfonts,amssymb,mathrsfs,amsmath,esint}
\usepackage{slashed, cancel}
\usepackage{framed}
\usepackage{mdframed}
\usepackage{simplewick} 
\allowdisplaybreaks
\usepackage[bottom]{footmisc}
\usepackage{geometry}

\usepackage{latexsym}
\usepackage{graphicx}
\graphicspath{{./Figures/}}
\usepackage[dvipsnames]{xcolor}
\usepackage{booktabs}
\usepackage{datetime}
\newdateformat{mydate}{\THEDAY{ }\monthname[\THEMONTH]{ }\THEYEAR}

\usepackage{tikz}
\usepackage{color}
\usepackage{framed}
\usepackage{hyperref}
\hypersetup{colorlinks, citecolor=bluscuro, linkcolor=black, urlcolor=bluscuro}
\definecolor{rossos}{cmyk}{0,1,1,0.55}
\definecolor{bluscuro}{rgb}{0.15, 0.2, .85}
\definecolor{bluchiaro}{cmyk}{1,.3,0.,0.1}

\graphicspath{{./Figures/}}

\newcommand{\be}{\begin{equation}}
\newcommand{\ee}{\end{equation}}
\newcommand{\bea}{\begin{eqnarray}}
\newcommand{\eea}{\end{eqnarray}}
\newcommand{\beq}{\begin{equation}}
\newcommand{\eeq}{\end{equation}}
\newcommand{\lp}{\left(}
\newcommand{\rp}{\right)}
\newcommand{\llp}{\left[}
\newcommand{\rrp}{\right]}
\def\beqa{\begin{eqnarray}}

\def\pbh{{\text{\tiny PBH}}}

\def\xipbh{\xi_\text{\tiny PBH}}

\def\npbh{\bar n_\text{\tiny PBH}}
\def\dpbh{\delta_\text{\tiny PBH}}
\def\mpbh{M_\text{\tiny PBH}}

\def\vr{{\vec{r}}}

\def\d{{\rm d}}

\def\eeqa{\end{eqnarray}}

\def\lsim{\mathrel{\rlap{\lower4pt\hbox{\hskip0.5pt$\sim$}}
    \raise1pt\hbox{$<$}}}         
\def\gsim{\mathrel{\rlap{\lower4pt\hbox{\hskip0.5pt$\sim$}}
    \raise1pt\hbox{$>$}}}         

\def\mpbh{M_\text{\tiny PBH}}

\def\xxi{r_\xi}

\def\msun{\ M_\odot}
\def\lsim{~\rlap{$<$}{\lower 1.0ex\hbox{$\sim$}}}
\def\bsim{~\rlap{$>$}{\lower 1.0ex\hbox{$\sim$}}}

\def\ln{{\rm ln}}

\usepackage[normalem]{ulem}

\newcommand{\arXiv}[2]{\href{http://arxiv.org/pdf/#1}{{\tt [#2/#1]}}}
\newcommand{\arXivold}[1]{\href{http://arxiv.org/pdf/#1}{{\tt [#1]}}}

\numberwithin{equation}{section}
\renewcommand\theequation{\arabic{section}.\arabic{equation}}

\usepackage{amsmath}
\usepackage{amsfonts}
\usepackage{amssymb}
\usepackage{graphicx, rotating}
\usepackage{epstopdf}

\usepackage{epsfig}
\usepackage{latexsym}
\usepackage{graphicx}
\usepackage{color}
\usepackage{amsmath,amssymb}
\usepackage{cite}
\usepackage{slashed}

\usepackage{hyperref}
\hypersetup{colorlinks, citecolor=bluscuro, linkcolor=black, urlcolor=bluscuro}
\definecolor{rossos}{cmyk}{0,1,1,0.55}
\definecolor{bluscuro}{rgb}{0.15, 0.2, .85}
\definecolor{bluchiaro}{cmyk}{1,.3,0.,0.1}

\numberwithin{equation}{section}

\makeatletter
\def\section{\@startsection {section}{1}{\z@}{-3.5ex plus -1ex minus
-.2ex}{2.3ex plus .2ex}{\large\bf}}
\def\subsection{\@startsection{subsection}{2}{\z@}{-3.25ex plus -1ex
minus -.2ex}{1.5ex plus .2ex}{\normalsize\bf}}
\makeatother
\newcommand{\captionfonts}{\small}
\makeatletter  
\long\def\@makecaption#1#2{%
 \vskip\abovecaptionskip
 \sbox\@tempboxa{{\captionfonts #1: #2}}%
 \ifdim \wd\@tempboxa >\hsize
   {\captionfonts #1: #2\par}
 \else
   \hbox to\hsize{\hfil\box\@tempboxa\hfil}%
 \fi
 \vskip\belowcaptionskip}
\makeatother   

\catcode`@=11
\def\marginnote#1{}
\newcount\hour
\newcount\minute
\newtoks\amorpm
\hour=\time\divide\hour
by60
\minute=\time{\multiply\hour by60 \global\advance\minute
by-\hour}
\edef\standardtime{{\ifnum\hour<12 \global\amorpm={am}
\else\global\amorpm={pm}\advance\hour by-12 \fi
\ifnum\hour=0
\hour=12 \fi
\number\hour:\ifnum\minute<10
0\fi\number\minute\the\amorpm}}
\edef\militarytime{\number\hour:\ifnum\minute<10
0\fi\number\minute}
\def\draftlabel#1{{\@bsphack\if@filesw
{\let\thepage\relax
\xdef\@gtempa{\write\@auxout{\string
\newlabel{#1}{{\@currentlabel}{\thepage}}}}}\@gtempa
\if@nobreak
\ifvmode\nobreak\fi\fi\fi\@esphack}
\gdef\@eqnlabel{#1}}
\def\@eqnlabel{}
\def\@vacuum{}
\def\draftmarginnote#1{\marginpar{\raggedright\scriptsize\tt#1}}
\def\draft{\oddsidemargin
0.0truein
\def\@oddfoot{\sl preliminary draft \hfil
\rm\thepage\hfil\sl\today\quad\militarytime}
\let\@evenfoot\@oddfoot
\overfullrule 3pt
\let\label=\draftlabel
\let\marginnote=\draftmarginnote
\def\@eqnnum{(\theequation)\rlap{\kern\marginparsep\tt\@eqnlabel}
\global\let\@eqnlabel\@vacuum}
}
\catcode`@=12
\def\dj{\hbox{d\kern-0.347em \vrule width 0.3em height 1.252ex depth
-1.21ex \kern 0.051em}}

\def\d{{\rm d}}
\def\ee{{\rm e}\,}

\def\ba{\bar a}

\def\Dirac{\,\raise.15ex\hbox{/}\mkern-13.5mu D}
\def\dirac{\,\raise.15ex\hbox{/}\kern-.57em \partial}
\def\aslash{\,\raise.15ex\hbox{/}\mkern-13.5mu A}
\def\shalf{{\ifinner {\textstyle \frac{1}{2}}\else \frac{1}{2} \fi}}
\def\sthreehalfs{{\ifinner {\textstyle \frac{3}{2}}\else \frac{3}{2} \fi}}
\def\sshalf{{\ifinner {\scriptstyle \frac{1}{2}}\else \frac{1}{2} \fi}}
\def\sfourth{{\ifinner {\textstyle \frac{1}{4}}\else frac{1}{4} \fi}}
\def\sphifour{{\ifinner {\textstyle \frac{1}{4!}}\else \frac{1}{4!} \fi}}
\def\lsim{\stackrel{<}{_\sim}}

\def\XXint#1#2#3{{\setbox0=\hbox{$#1{#2#3}{\int}$}
    \vcenter{\hbox{$#2#3$}}\kern-.5\wd0}}

\def\bea{\begin{eqnarray}} \def\eea{\end{eqnarray}}
\def\be{\begin{eqnarray}} \def\ee{\end{eqnarray}} \def\nn{\nonumber}

\newcommand{\promille}{%
 \relax\ifmmode\promillezeichen
       \else\leavevmode\(\mathsurround=0pt\promillezeichen\)\fi}
\newcommand{\promillezeichen}{%
 \kern-.05em%
 \raise.5ex\hbox{\the\scriptfont0 0}%
 \kern-.15em/\kern-.15em%
 \lower.25ex\hbox{\the\scriptfont0 00}}

\def\d{{\rm d}}
\def\nn{\nonumber}

\def\cs2{c_{s}^{2}}

 \def\be   {\begin{equation}}   \def\ee   {\end{equation}}
 \def\ba   {\begin{array}}      \def\ea   {\end{array}}
 \def\bea  {\begin{eqnarray}}   \def\eea  {\end{eqnarray}}
 \def\bean {\begin{eqnarray*}}  \def\eean {\end{eqnarray*}}

\begin{document}
\parskip=10pt
\baselineskip=18pt

{\footnotesize  }
\vspace{5mm}
\vspace{0.5cm}

\begin{center}

\def\thefootnote{\fnsymbol{footnote}}

\topskip=70pt

{ \large 
\bf Primordial Black Holes from  Broad Spectra:   Abundance and  Clustering
}
\\[1.2cm]
{ Azadeh Moradinezhad Dizgah $^{1,2}$, Gabriele Franciolini$^{1}$,  and Antonio Riotto$^{1,3}$} \\[0.6cm]

{\small \textit{$^1$ D\'epartement de Physique Th\'eorique and Centre for Astroparticle Physics (CAP), \\
Universit\'e de Gen\`eve, 24 quai E. Ansermet, CH-1211 Geneva, Switzerland}}

{\small \textit{$^2$ Department of Physics, Harvard University, 17 Oxford Street, Cambridge, MA 02138, USA}}  

{\small \textit{$^3$ CERN,
Theoretical Physics Department, Geneva, Switzerland}}

\vspace{.2cm}

\end{center}

\vspace{.8cm}

\begin{center}
\textbf{Abstract}
\end{center}
\noindent
A common mechanism to form  primordial black holes in the early universe is by enhancing at small-scales the scalar perturbations generated during inflation. If these fluctuations have a large enough amplitude, they may collapse into primordial black holes upon horizon re-entry. Such primordial black holes may comprise the totality of the dark matter. We offer some considerations about the formation and clustering of primordial black holes when the scalar perturbations are characterised by a broad spectrum. Using the excursion set method, as well as the supreme statistics, we show  that the cloud-in-cloud phenomenon,  for which  small mass primordial black holes may be  absorbed by bigger mass ones, is basically absent. This is due to the fact that the formation of a primordial black hole is an extremely rare event. We also show that, from the point of view of mass distribution, broad and narrow spectra give similar results in the sense that the mass distribution is tilted towards a single mass. Furthermore, we argue that primordial black holes from broad spectra are not clustered at formation, their distribution is  dominantly Poissonian.

\vspace{.5in}

\def\thefootnote{\arabic{footnote}}
\setcounter{footnote}{0}
\pagestyle{empty}

{\let\thefootnote\relax \footnote{{\small Email:  Azadeh.MoradinezhadDizgah@unige.ch, Gabriele.Franciolini@unige.ch, Antonio.Riotto@unige.ch}}}

\newpage
\pagestyle{plain}
\setcounter{page}{1}

\section{Introduction}
\noindent
The  detection of gravitational waves  produced by the merger of two $\sim 30\, M_\odot$ black holes \cite{ligo1} has started the era of  gravitational wave astronomy.  At the same time, it has revived the interest in the physics of Primordial Black Holes (PBHs) and the exciting hypothesis that all (or a significant part) of the dark matter of the universe might be  in the form of PBHs \cite{bird} (see Ref. \cite{revPBH} for a review and additional references). 

PBHs in the early universe might owe their origin to the small-scale enhancement of the primordial curvature perturbation $\zeta$ generated during inflation \cite{s1,s2,s3}. After inflation, during reheating, these small-scale perturbations are communicated to the radiation fluid.  PBHs  (of mass roughly equal to the horizon mass) are generated at horizon re-entry, if the density perturbation  $\delta$ has a large enough amplitude, and hence the radiation density contrast is larger than a critical value $\delta_{c}$ \cite{musco}.

The abundance of PBHs is estimated using either the peak theory \cite{bbks} or the Press-Schechter (PS) approach \cite{PS}. Assuming spherical symmetry, to determine the characteristic scale and amplitude of the fluctuations that collapse to form a PBH, one defines the compaction function $C(r)$ as the ratio of mass-excess within a sphere of radius $r$ to the areal radius at $r$,   \cite{compac, musco}
\be
C(r) = \frac{2\left[M(r, t)-M_b(r, t)\right]}{R(r, t)}.
\ee
Here $R(r,t)$ is the areal radius, $M(r,t)$ is the mass within a sphere of radius $r$ centred on the peak location, and $M_b(r,t)$ is the background mass within the same radius, calculated with respect to a flat FRW Universe. The criterion for collapse widely used in the recent literature states that a PBH forms if the maximum of the compaction function is above a certain threshold, found numerically \cite{musco}. It can be shown that the amplitude of the fluctuations in terms of the excess mass within a spherical volume is equivalent to the local value of compaction function \cite{musco}. Therefore the scale of fluctuations relevant for the formation of PBHs, commonly referred to as $r_m$, can be determined by maximizing the compaction function. Given that the PBHs form from large-amplitude fluctuations, one can apply the peak statistics \cite{bbks}, assuming spherical peaks (which is a good approximations since the PBHs are rare events), to calculate $r_m$. 

Considering the average shape around a peak of height $\delta_0$ \cite{hof}
\be\label{av0}
\overline{\delta}(r)=\delta_0\frac{\xi(r)}{\sigma^2},
\ee
with $\xi(r)$ being the two-point correlator in real space if working in the comoving slicing, 
\be
\label{av}
\xi(r)=
\int\frac{{\rm d} k}{k}\frac{\sin k r}{k r}{\cal P}_{\delta}(k,t), \quad {\cal P}_{\delta}(k,t)=\frac{16}{81}\left(\frac{k}{aH}\right)^4{\cal P}_{\zeta}(k),
\ee
the threshold for the formation of a PBH is defined in terms of volume-averaged mass-excess, which is three times the peak profile at $r_m$ \cite{musco}
\be
\delta_c=3\overline{\delta}(r_m).
\ee
The corresponding critical peaks amplitude, $\delta_0$, has to be found numerically. 

Once the critical threshold is known, the abundance of PBHs can be computed using threshold statistics. For Gaussian fluctuations\footnote{For the non-Gaussian extension, see \cite{Young:2013oia,Young:2015cyn,ng1,Yoo:2018esr,Atal:2018neu,ng2,ng3,ng4,Atal:2019cdz,muscong,Yoo:2019pma}. The non-Gaussian corrections are going to be neglected in this paper and the study of their impact on the present discussion is left for future work.}, the mass fraction of PBHs at the time of collapse, $\beta$, is given by 
\be
\beta(M(r_{m}))=\int_{\delta_c}\frac{\d\delta_m }{\sqrt{2\pi}\sigma_m}\,e^{-\delta_m^2/2\sigma^2_m}\simeq 6 \cdot 10^{-9}\left(\frac{M(r_{m})}{M_\odot}\right)^{1/2},
\ee
where in the last passage we have assumed that the PBHs contribute to the totality of the dark matter, and  the  volume-averaged field and its variance are given by
\begin{eqnarray}
\label{con}
\delta_m&=&\frac{3}{r_m}(aH)^2\int_0^{r_m}{\rm d}r\, r^2\, \delta(r,t),\nonumber\\
 \sigma^2_m&=&\int\frac{{\rm d} k}{k}{\cal P}_{\delta_m}(k)T^2(k,r_m),
\end{eqnarray}
with $T(k,r_m)$ being the radiation transfer function
\begin{equation}
\label{tf}
T(k,r_m)=3\frac{\sin (kr_m/\sqrt{3})-(kr_m/\sqrt{3})\cos(kr_m/\sqrt{3})}{(kr_m/\sqrt{3})^3}.
\end{equation}
We have conveniently multiplied the smoothed overdensity by the the factor $(a H r_m)^2$ to obtain a time-independent quantity  which can be thus calculated on superhorizon scales \cite{musco,young}.

The dynamics of PBH formation, the PBH mass and mass function are rather well understood in the case of a narrow curvature perturbation power spectrum, where only one length scale $1/ k_\star$  dominates. A benchmark power spectrum often used in the literature is the  Dirac-delta type 
\be
\label{power}
{\cal P}_\zeta ={\cal P}_0  \,\delta_D(\ln \,k- \ln\, k_\star).
\ee
 In such a case the scale $r_m$ is the order of $1/k_\star$ \cite{muscogermani} and PBHs are formed with a mass $\sim M(1/k_\star)$.

For broad spectra  things are  more blurred. Consider a power spectrum which extends over a large range of scales, say from $k_l$ to $k_s\gg k_l$. In a  realistic universe there are  many perturbations superimposed on top of each other with varying scales, amplitudes, and physical locations (i.e. different Fourier modes with varying wavelength, amplitude and phase). One therefore expects peaks with various amplitude and, above all, with different length scales, running from $\sim 1/k_l$ to $\sim 1/k_s$. Each peak profile will have its own scale $r_m$ associated with it. 

Several questions will then arise; What is the distribution of PBH masses for broad spectra? Are the small PBHs swallowed by bigger PBHs, i.e. is there a so-called cloud-in-cloud problem, similar to that for the dark matter halo formation?  Are PBHs clustered differently from a Poisson distribution, contrary to what happens for a narrow power spectrum  \cite{v}? The goal of the paper is to offer some considerations about these questions. Our basic findings are the following. 

Using the excursion set method (with further support from an alternative technique based on the supreme statistics), we will show that the cloud-in-cloud phenomenon is basically absent for PBHs. This is due to the fact that the probability of forming a PBH (if we wish them to compose all or most of the dark matter) is very small, i.e. PBHs are very rare events 
\footnote{If PBHs constitute only a small fraction of the dark matter, their mass fraction is smaller, and hence they are even rarer events. Therefore, our results hold even stronger.}. This in practice prevents pre-existing PBHs to be swallowed by bigger ones. Also, despite the broadness of the power spectrum of the curvature perturbation, the production of PBHs is tilted towards either small-scale (small mass) or large-scale (large mass) PBHs. In this sense, the mass spectrum of PBHs does not differ from the one derived for a narrow power spectrum. This result also suggests that the clustering properties at formation will be similar for broad and narrow spectra: PBHs at their birth are likely to be distributed randomly following a Poisson distribution. 

The paper is organized as follows. In section 2 we will discuss the threshold for formation of PBHs in the case of a broad spectrum. Section 3 is a brief summary of the excursion set method and its consequences when applied to the physics of PBHs formation. Section 4 supports some of the results found in the previous section through the so-called supreme statistics. Section 5 is devoted to the clustering, and section 6 contains our conclusions.



\section{Broad spectra and  their threshold}
\setcounter{section}{2}
\noindent
The first question to ask for broad spectra is what is the critical threshold. 
Here we follow Ref. \cite{young}  and  imagine to separate the overdensity into long- (background) and short-modes  (foreground)
\be
\delta(\vec x,t)=\delta_l(\vec x,t)+\delta_s(\vec x,t)=\int_{k_l}^{k_{\rm ref}}\frac{\d^3 k}{\lp 2 \pi \rp^3}\,\delta_{\vec k }\,e^{-i\vec k\cdot \vec x}+\int_{k_{\rm ref}}^{k_s}\,\frac{\d^3 k}{\lp 2 \pi \rp^3}\,\delta_{\vec k }\,e^{-i\vec k\cdot \vec x}, \quad 
k_l\ll k_{\rm ref}\ll k_s,
\ee
where, for simplicity, we suppose that both the long and short mode power spectra are flat, even though they can have different amplitudes. The operation of selecting  the longer modes can be in principle done by a smoothing\footnote{One should not give too much emphasis on the use of the smoothing to  determine the average profile shape. The physical evolution of a perturbation depends on the profile shape, but it should not  depend on the smoothing function. The calculation of the abundance does depend on the choice of window function, but this just highlitghts the  uncertainty in the calculation rather than suggesting different amounts of PBHs form \cite{young}.} over  volumes of radius $\sim 1/k_{\rm ref}$.  Since what matters is always the smallest length scale \cite{musco}, computing the  average profile of the short-mode perturbations delivers a value $r_{m,s}$ which is approximately given by $3.5/k_s$ \cite{muscogermani}. Clearly the long-mode perturbations have a much larger characteristic scale $r_{m,l}\sim 3.5/k_{\rm ref}$. 

\begin{figure}[t!]
	\centering
	\includegraphics[width=.495\textwidth]{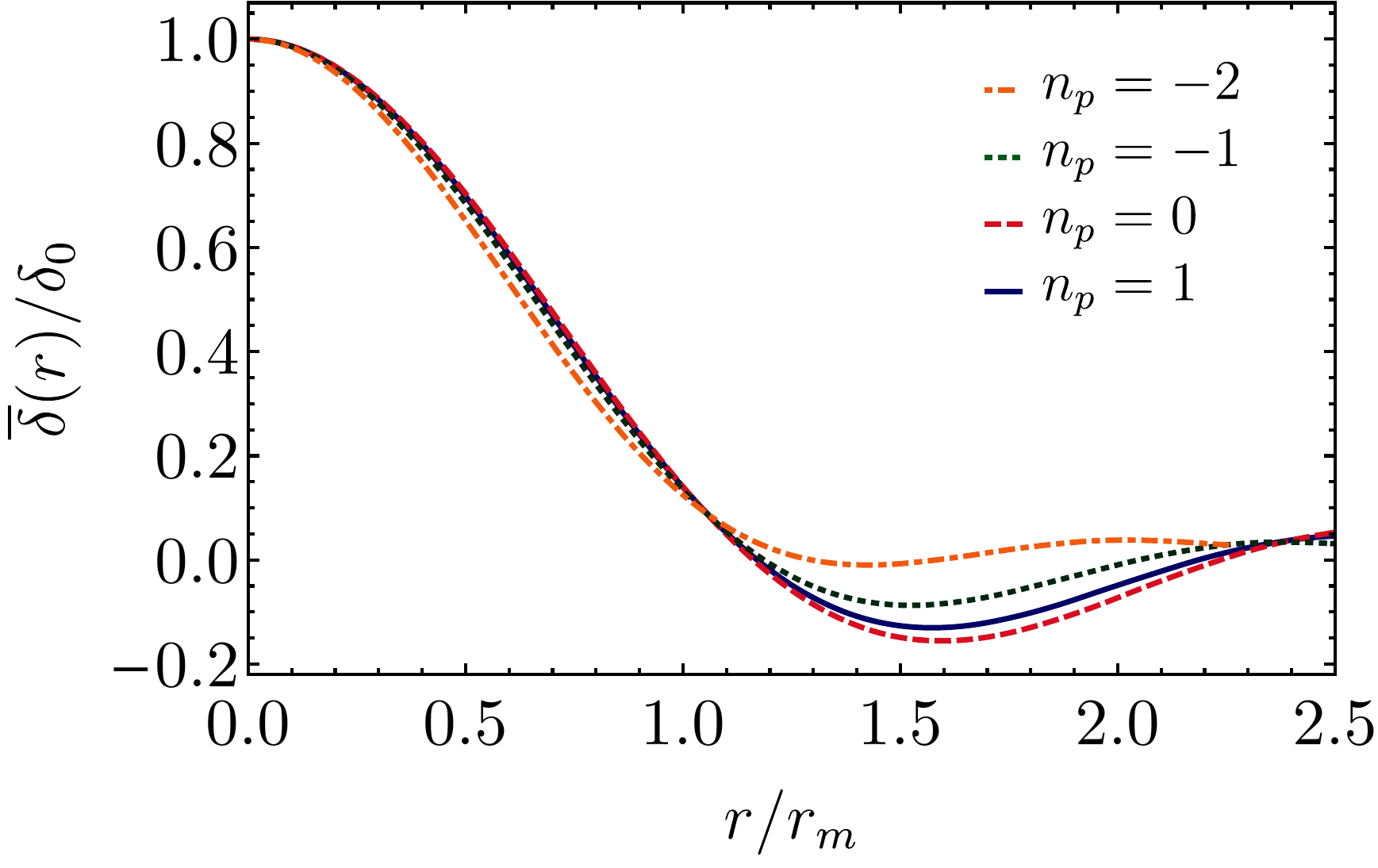}
		\caption{\it The peak profiles for the power spectrum given in Eq. \eqref{eq:broad_spec}, for different values of $n_p$.}
		\label{profile}
		\end{figure}

Although the values of $r_m$ for short and long mode differ, it is important to note that the critical threshold value is the same for both cases and is $\delta_c\simeq 0.51$.  In figure \ref{profile} we show the typical profile in Eq.~\eqref{av0}, for a generalised broad power spectrum of the form 
\be\label{eq:broad_specc}
{\cal P}_\zeta \approx{\cal P}_0  (k/k_{s})^{n_p}\Theta \left ( k_{s}-k\right) \Theta \left ( k-k_{l}\right),
\ee
where $\Theta$ is the Heaviside step function,  $k_s=10^4 \,k_l$ and $n_p=-2,-1,0,1$. The approximate sign in Eq. (\ref{eq:broad_specc}) reminds us that transition from 0 to $k_l$ and from $k_s$ from 0 is not sudden in real cases. 
\noindent
%
\noindent
The plot shows that the profiles are nearly identical within their corresponding distance $r_m$ from the centre, which  leads to the same critical threshold (see also \cite{young}).  We conclude that for broad spectra, peaks with different values of $r_m$  will eventually collapse if the volume-averaged over a sphere of radius $r_m$ overcomes the same threshold $\delta_c$. 

Notice that causality prevents the  large-scale perturbations longer than the Hubble horizon from affecting the formation of light PBH on short-scales. On the other side such PBHs formed at small-scales do not affect the fluctuations on large-scales after they re-enter the horizon. Indeed, and more in general,  the effect of a set of self-gravitating particles in a virialized cluster does not alter the behaviour of the perturbations on much larger scales. In this sense, gravitational interactions in the cosmological setting are renormalizable.

\section{The excursion set method and  PBHs}
\noindent
Having argued that peaks with different characteristic scales $r_m$ have the same threshold, the next question is what are the masses  which count as far as the PBH formation is concerned.

As already mentioned in the introduction, the abundance of PBHs can be determined by the Press-Schecter method, very much as done when  computing the mass function  of dark matter halos. An extension of it, the so-called excursion set theory \cite{Bond}, corrects  the original argument of Press and Schechter  which miscounts the number of virialized dark matter objects because of the so-called cloud-in-cloud problem. In the dark matter spherical collapse model one assumes that  a region of radius $R$, with a smoothed  density contrast $\delta_R$, collapses and virializes once $\delta_R$ exceeds a critical value. This procedure misses the cases in which, on a given smoothing scale $R$,  $\delta_R$ is below the threshold, but still it happened to be above the threshold at some scale  $R'>R$. Such a configuration corresponds to a virialized object of mass $M'>M$.  

In the excursion set method the density perturbations evolve stochastically with the smoothing scale, and the problem of computing the probability of halo formation is mapped into the so-called first-passage time problem in the presence of a barrier. The cloud-in-cloud problem is avoided by counting only the trajectories which pass the threshold for the first-time.

The excursion set method is therefore well-suited to develop a better understanding of formation of PBHs form a broad spectrum, above all the cloud-in-cloud problem. For the reader not familiar with the excursion set theory, we provide a brief overview of the method.  The expert reader might wish to skip this part and go directly to section \ref{bspec}.

\subsection{A brief review of the excursion set method}
\noindent
Consider the density contrast $\delta({\vec x})$, smoothed on a particular scale $R$ as
\be\label{dfilter}
\delta({\vec x},R) =\int \d^3x'\,  W(|{\vec x}-{\vec x}'|,R)\, \delta({\vec x}'),
\ee
where $W(|{\vec x}-{\vec x}'|,R)$ is the window function. Since the convolution in real space is a product in Fourier space, the Fourier transform of the smoothed field is 
\be
\delta(\vec k,R) = \widetilde{W}({\vec k},R) \delta(\vec k)
\ee
with $\widetilde{W}({\vec k},R)$ being the Fourier transform of the real-space window function. There are various choice for the window function. As we discuss below, for a sharp filter in $k$-space, defined as
\be\label{Wsharpk}
\widetilde{W}_{{\rm sharp}-k}(k,k_f)=\Theta(k_f-k),
\ee
with $k_f=1/R$, $k=|{\vec k}|$, and $\Theta$ being the Heaviside step function, the formulation of the excursion-set theory is the simplest. 

In Eq. \eqref{dfilter}, the density field is independent of time. We will denote its variance as $S$ in the following (while $\sigma_R^2$ is reserved to the time-dependent variance). This is because in the framework of excursion set theory, to determine the collapsed fraction, one determines the evolution of the density field in terms of the variance at fixed time. The formation history of the collapsed objects, can be studied by retrieving the time-dependance via promoting the collapse threshold to be time-dependant, while keeping the variance fixed in time. This is due to the fact that the probability of forming virialized objects is determined in terms of the ratio $\delta_c/\sigma_R$, with $\sigma_R$ being the square root of the variance of the smoothed density field. We will further illustrate on this for the particular case of PBHs at the end of this section.  

One proceeds by studying the ``evolution" of $\delta({\vec x},R)$ with $R$, at a fixed
value of ${\vec x}$, which is given by
\be\label{padeta1}
\frac{\partial\delta_R}{\partial R} =\zeta (R),
\ee
where
\be\label{zetadelta}
\zeta(R)\equiv \int \frac{\d^3k}{(2\pi)^3}\,
\widetilde{\delta}({\vec k} ) \frac{\partial\tilde{W}(k,R)}{\partial R}.
\ee
Without loss of generality, we set $\vec x = 0$, and to keep the notation compact, denote the smoothed field at $\vec x=0$ as $\delta_R$. Since the modes $\widetilde{\delta}({\vec k} )$ are stochastic variables, $\zeta(R)$ is also a stochastic variable, and Eq. (\ref{padeta1}) has the form of a Langevin equation, with $R$ playing the role of time, and
$\zeta(R)$ playing the role of noise. Statistical properties of a Gaussian distributed $\delta_R$, and hence $\zeta(R)$, are fully determined in terms of their connected two-point correlator, which for $\zeta(R)$ is given by
\be\label{corr2eta}
\langle \zeta(R_1)\zeta(R_2)\rangle =
\int_{-\infty}^{\infty}\d \ln k\,\, \Delta^2(k)
\frac{\partial\widetilde{W}(k,R_1)}{\partial R_1}
\frac{\partial\widetilde{W}(k,R_2)}{\partial R_2},
\ee
where $\Delta^2(k)=k^3P_\delta(k)/(2\pi^2)$.  For a generic filter, the right-hand side is a  function of  $R_1$ and $R_2$, different from a Dirac-delta 
$\delta_D(R_1-R_2)$. 

For the particular case of sharp $k$-space filter, things simplify considerably, and one retrieves the Langevin equation with Dirac-delta noise. This can be clearly seen using $k_f=1/R$ instead of $R$, and defining $Q (k_f)=-(1/k_f) \zeta(k_f)$ to get
\be
\label{padeta2}
\frac{\partial\delta(k_f)}{\partial \,\ln \,k_f} =Q (k_f),
\ee
with
\be\label{corr2etab}
\langle Q ({k_f}_1)Q ({k_f}_2)\rangle =
\Delta^2({k_f}_1) \delta_D(\ln {k_f}_1-\ln {k_f}_2).
\ee
The equations can be further simplified using the variance $S$ of the overdensity as a ``pseudotime''
variable, 
\be\label{mu2RW2}
S(R) =\int_{-\infty}^{\infty} \d \,\ln\, k \, \Delta^2(k) \, |\widetilde{W}(k,R)|^2,
\ee
which for a sharp $k$-space filter becomes
\be
S(k_f) =\int_{-\infty}^{\ln \,k_f} \d\,\ln\, k\, \Delta^2(k).
\ee

Finally, redefining $\eta (k_F) =Q(k_F)/\Delta^2(k_F)$, Eq. \eqref{padeta2} can be re-written as 
\be\label{Langevin1}
\frac{\partial\delta(S)}{\partial S} = \eta(S),
\ee
with
\be\label{Langevin2}
\langle \eta(S_1)\eta(S_2)\rangle =\delta_D (S_1-S_2).
\ee

\begin{figure}[t!]
	\centering
	\includegraphics[width=.49\textwidth]{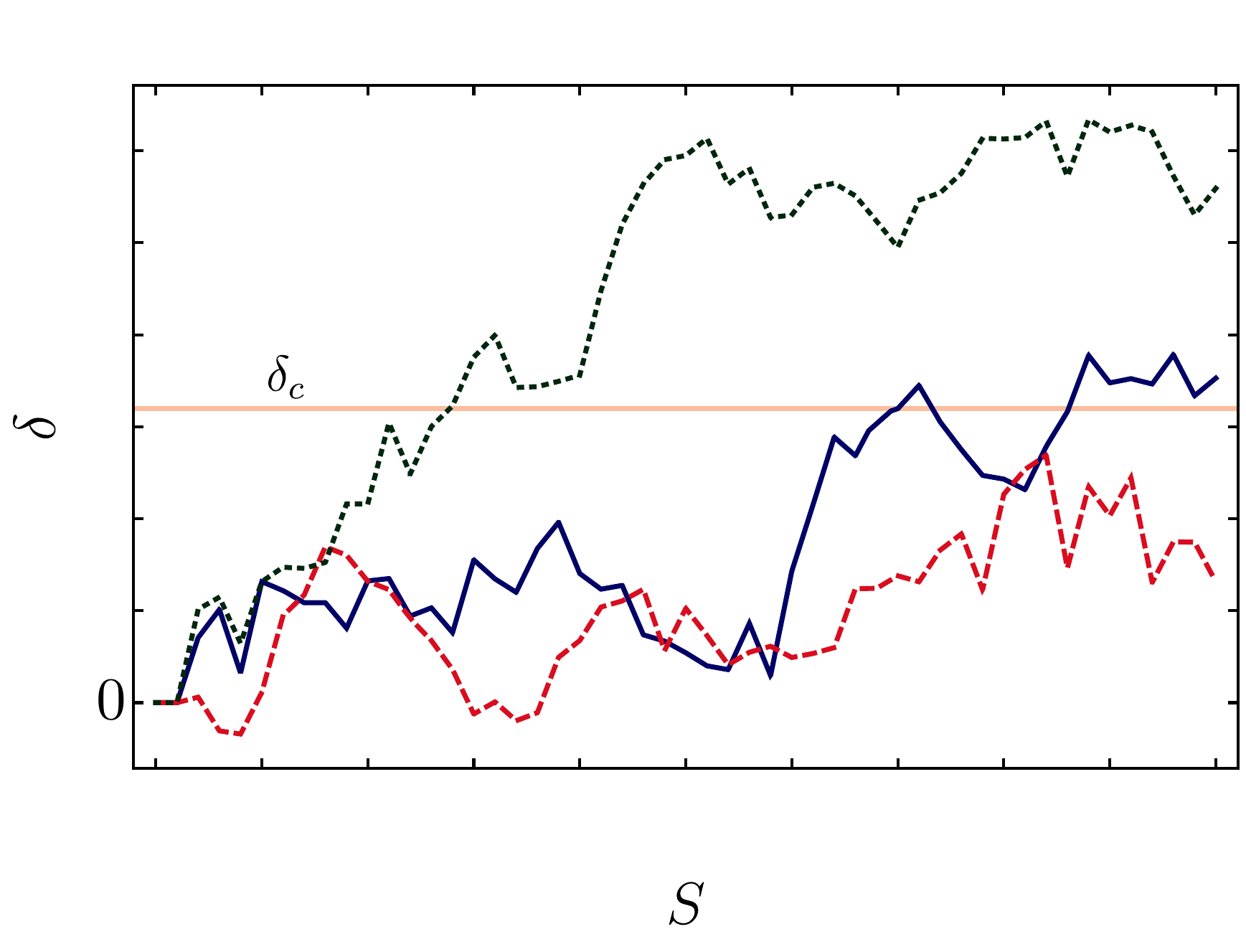}
		\caption{\it A schematic picture of the random walk performed by the smoothed overdensity.} 
		\label{exc}
\end{figure}

Therefore, the evolution of the density field in terms of the variance (playing the role of time) obeys the Langevin equation with a Gaussian white noise. This implies that $\delta(S)$ performs a Brownian random walk, with respect to the ``time'' variable $S$, with no memory from one step to the other\footnote{If the window function is not a step function in momentum space, and is, for instance, a step function in real space (which gives rise to the volume-averaged field in Eq. (\ref{con})), the random walk of the smoothed overdensity is non-markovian, introducing memory effects from one step to the other (the white-noise case is markovian and no memory effects are present). This case has been studied in Refs. \cite{MR1,MR2,MR3} using the path-integral methods, but is much more involved than the case of a white-noise. For the sake of simplicity,  in this paper we restrict ourselves to  the markovian case originally studied in Ref.~\cite{Bond}.}. Following Ref. \cite{Bond}, we refer to the evolution of $\delta$ as a function of $S$ as a ``trajectory'' (see a schematic picture in figure \ref{exc}). 

It is well-known that for a stochastic variable evolving according to Langevin equation with white noise, the distribution obeys the Fokker-Planck  equation. Therefore, the distribution of $\delta_R$  can be determined from
\be\label{FPdS}
\frac{\partial P}{\partial S}=\frac{1}{2}\, \frac{\partial^2 P}{\partial \delta^2}.
\ee
The Press-Schecter approach is equivalent to solving this equation, setting the boundary condition that the distribution vanishes at $\delta \rightarrow \pm \infty$, and accounting for all the trajectories that exceed the threshold of collapse $\delta_c$. This leads to the so-called ``cloud-in-cloud'' problem, which is associated with trajectories that make multiple
crossings of the  threshold. This problem is cured in excursion set theory approach by accounting for the {\it first} up-crossing of the trajectories. This is equivalent to calculating  the lowest value of $S$ (or, equivalently, the highest value of $R$) for which the trajectory pierces the threshold. Once a trajectory passes the threshold, a virialized object forms,  and this trajectory should be excluded from further consideration. Therefore, the solution of the cloud-in-cloud problem is implemented by
imposing the boundary condition
\be
\left. P (\delta,S)\right|_{\delta=\delta_c}=0, 
\ee
for which the solution of the FP equation is given by
\be\label{PiChandra}
P (\delta,S)=\frac{1}{\sqrt{2\pi S}}\,
\left[  e^{-\delta^2/(2S)}- e^{-(2\delta_c-\delta)^2/(2S)} \right].
\ee
The fraction $F(S)$ of trajectories that have crossed the threshold at
``time'' smaller or equal to $S$ is then calculated from the fraction of
trajectories that at ``time'' $S$ have never crossed the threshold, so
\be\label{F(S)1menoint}
P(\delta>\delta_c)=1-\int_{-\infty}^{\delta_c}\d\delta\, P(\delta,S)={\rm Erfc}\left(\frac{\delta_c}{\sqrt{2S}}\right),
\ee
which  is twice the Press-Schecter prediction.

Excursion set theory can also be used to study the formation history of virialized objects. For dark matter halos, this was briefly discussed in the classic paper of Bond {\it et al.}, \cite{Bond} and further elaborated in Ref. \cite{LC}. As we briefly discussed earlier, the time evolution of the density field is conveniently recovered  by absorbing the time-dependance into the collapse threshold, and defining a time-dependant threshold $\omega(a) \equiv  \delta_c/D(a)$, with $D(a)$ being the linear growth factor. Therefore, the collapse is determined in terms of the linear density field evolved to the time of interest. Since for PBHs, the threshold scales like $\omega(a) \equiv \delta_c/a^2$, the barriers at two different time are related to one another as 
\be\label{rt}
\omega(a_1) = \left(\frac{a_2}{a_1}\right)^2 \omega(a_2).
\ee
In other words, the barrier is lowered as time passes, while the variance is constant. In this way, the real time evolution is given by the time evolution of the barrier, which for PBHs is in one-to-one correspondence with the increase of the comoving Hubble length. 

\begin{figure}[t!]
	\centering
	\includegraphics[width=.49\textwidth]{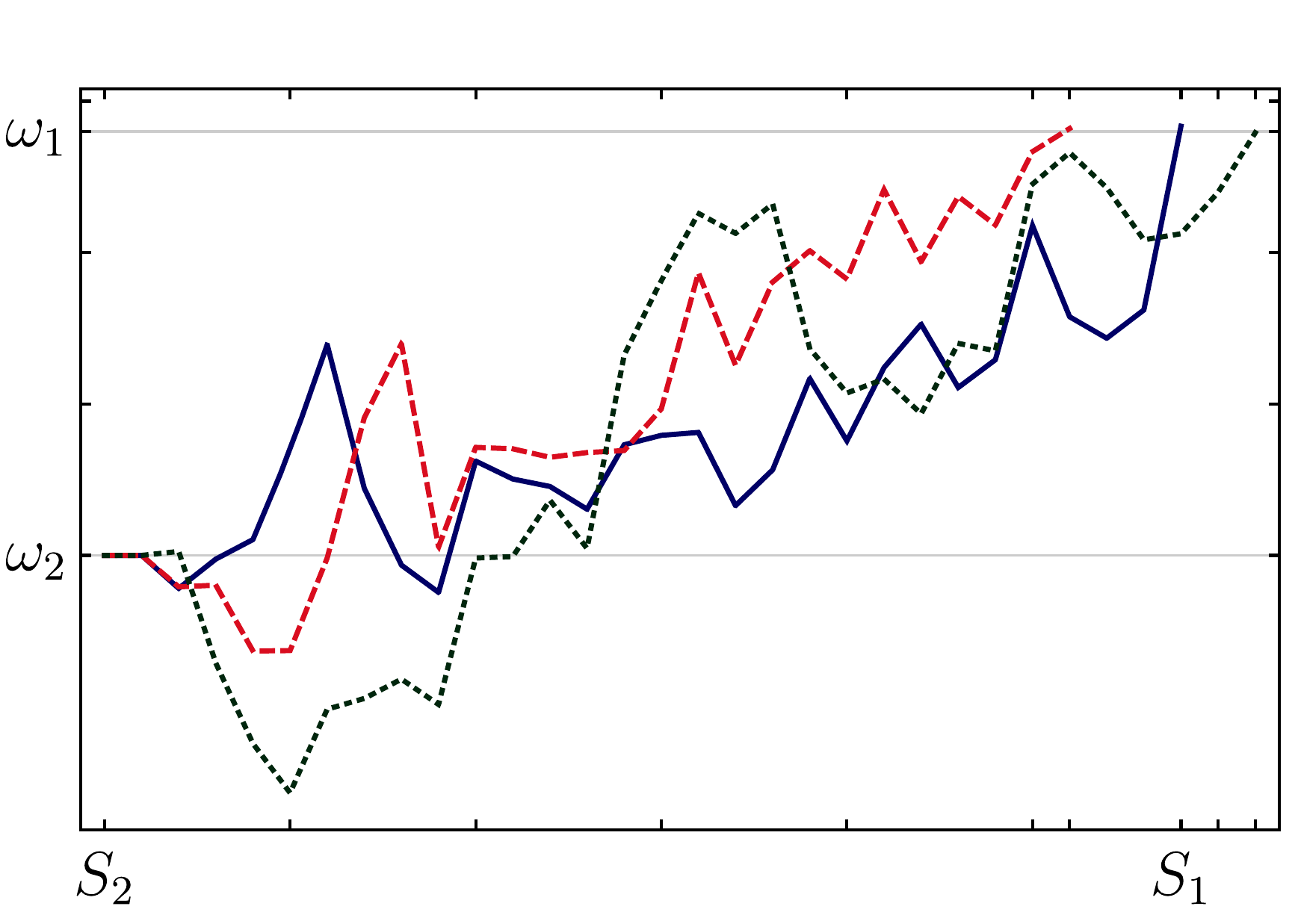}
		\caption{\it A schematic picture of the random walk performed by the smoothed overdensity with a two-barrier problem.} 
		\label{exc-2}
\end{figure}

A great deal of insight into the formation history can be gained by considering the ``two-barrier'' problem \cite{LC}, which allows us not only to calculate the probability to form a virialized object at a given time, but also the conditional probability that a trajectory with a first up-crossing of the barrier $\omega_1$ at some $S_1$ will have a first upcrossing of $\omega_2$ between $S_2$ and $S_2+\d S_2$ with $S_1\gg S_2$ and $\omega_1>\omega_2$ (see figure \ref{exc-2} for a schematic picture). This conditional probability is given by
\begin{eqnarray}
\label{fund}
P(S_2,\omega_2|S_1,\omega_1)\d S_2&=&\frac{P(S_1,\omega_1|S_2,\omega_2)}{P(S_1,\omega_1)}\d S_2\nonumber\\
&=&\frac{1}{\sqrt{2\pi}}\left[\frac{S_1}{S_2(S_1-S_2)}\right]^{3/2}\frac{\omega_2(\omega_1-\omega_2)}{\omega_1}\exp\left[
-\frac{(\omega_2 S_1-\omega_1 S_2)^2}{2S_1 S_2(S_1-S_2)}\right]
\d S_2.
\end{eqnarray} 

For PBHs, analogous to halos, the trajectories provide the variation of $\delta_R$ at fixed time. However, the condition for forming a PBH as well as its mass apply when a given scale crosses the horizon. One must generate the trajectories at a given fixed time and then evolve each value of $\delta_R$ backwards or forwards in time till a given scale of interest crosses the horizon \cite{lg}. Ultimately one is interested in the mass $M_H$ associated with a given scale when the latter  enters the horizon, $R\sim 1/aH\sim a\sim M_H^{1/2}$ (in the radiation-dominated case), so that, still at fixed time
\be
\label{aa}
S(R_1)
=S(R_2) \left(\frac{M_{H_1}}{M_{H_2}}\right)^{-(4+n_p)/2}.
\ee 
Eq. \eqref{fund}  then gives the  conditional probability that a PBH  of mass $M_1$ present at the time $t_1$ has merged into a PBH of mass between $M_2$ and $M_2+\d M_2$ at time $t_2> t_1$. This quantity therefore seems the most suitable one to discuss the fate of the small PBHs in the broad spectrum case.


\noindent
\subsection{Broad spectra with a blue tilt}\label{bspec}
\noindent
Having reviewed basics of the excursion set theory,  let us get back to the questions of what are the relevant scales for the formation of PBHs for a broad primordial spectrum. We advocate here that for realistic power spectra with a blue tilt
\be\label{eq:broad_spec}
{\cal P}_\zeta \approx {\cal P}_0  (k/k_{s})^{n_p}\Theta \left ( k_{s}-k\right) \Theta \left ( k-k_{l}\right),\quad n_p>0,
\ee
it  is always the smallest scale (the largest momentum) that is relevant. In other words, if a broad spectrum is characterized by a range of scales running from $1/k_l$ to $1/k_s$, production of PBHs will be dominated by PBHs with mass $M(r_{m,s})$. To show this we can resort to two alternative techniques, one based on the excursion set that we discuss in this section, and the other on the so-called supreme statistics to which we devote the next section.

Let us consider the fluctuations on small and large scales which collapse to form PBHs at the time when the scale factors are $a_s$  and $a_l$ correspondingly.  From what we discussed previously, the critical threshold for formation, $\delta_c$, is the same for all the PBHs. This means that the rescaled thresholds in Eq. \eqref{rt} are hierarchical. Neglecting the radiation transfer function, the variances (at fixed time) as a function of the scale are also hierarchical 
\begin{eqnarray}
\label{conds}
S_s &\simeq& \frac{16}{81(n_p+4)}{\cal P}_0k_s^{4},\nonumber\\
S_l &\simeq& \frac{16}{81(n_p+4)}{\cal P}_0\left(\frac{k_l}{k_s}\right)^{n_p}k_l^4\ll S_s.
\end{eqnarray}
From the excursion point of view, the separation of scales $k_s\gg k_l$ with a blue spectrum is therefore  given by the hierarchies
\be
\omega(a_s)\gg \omega(a_l)\quad {\rm and} \quad S_s\gg S_l
\ee
and
\be
\label{ooo}
\sqrt{\frac{S_s}{S_l}}=\left(\frac{k_s}{k_l}\right)^{(4+n_p)/2}\gg \frac{\omega(a_s)}{\omega(a_l)}=\left(\frac{k_s}{k_l}\right)^{2}.
\ee
Notice that the last condition implies that small PBHs are generated with larger abundances than the big ones, since the abundances are given by
\bea
\beta(M(r_{m,l}))&\sim& e^{-\omega^2(a_l)/(2S_l)} \nonumber \\
\beta(M(r_{m,s})) &\sim& e^{-\omega^2(a_s)/(2S_s)}
\eea
and hence,
\be
\beta(M(r_{m,l})) \ll   \beta(M(r_{m,s})).
\ee
Taking the above limit in Eq. (\ref{fund}) we have
\begin{align}
P(S_l,\omega(a_l)|S_s,\omega(a_s))\d S_l
&\simeq \frac{\omega(a_l)}{\sqrt{2\pi}\, S_l^{3/2}}\, e^{-\omega^2(a_l)/(2S_l)}\d S_l
\nn \\
&=\frac{1}{2} \beta(M(r_{m,l})) \frac{\omega ^2(a_l)}{S_l^2} \d S_l .
\end{align}
Therefore, when larger modes re-enter the horizon, PBHs of higher masses form with a negligible abundance. Moreover, even if they do form, they  do not engulf the previously formed lighter PBHs. In other words, the cloud-in-cloud phenomenon does not exist in practice for broad spectra with a blue tilt. 

The final spectrum of masses is dominated by the small PBHs. One could argue that when longer modes enter the horizon, the PBHs could  also form if $\sigma_l$ is sufficiently large.  However, since at each horizon crossing time $\sigma_s\gg\sigma_l$, this implies that the variance of the small-scales would be dangerously in the non-perturbative regime. In this sense broad spectra with a blue tilt give predictions similar to the ones of narrow spectra.

\subsection{Flat broad spectra}
\noindent
Next, let us consider a broad spectrum which is flat, 
\be\label{eq:broad_flat}
{\cal P}_\zeta \approx {\cal P}_0  \Theta \left ( k_{s}-k\right) \Theta \left ( k-k_{l}\right).
\ee
In such a case, the hierarchies for the thresholds and the variances is maintained
\be
\omega(a_s)\gg \omega(a_l)\quad {\rm and} \quad S_s\gg S_l.
\ee
However (see Eq. (\ref{ooo}) with $n_p=0$)
\be
\sqrt{\frac{S_s}{S_l}}=  \frac{\omega(a_s)}{\omega(a_l)}.
\ee
This condition implies that small PBHs are generated with the same abundance as large ones. 

The probability that a small PBHs have been incorporated into a new halo of a larger mass at  $S(<S_s)$ at later times is given by \cite{LC} (we approximate the
lower limit of integration to zero, $S_l \simeq 0$) 
\begin{eqnarray}
P(S,\omega(a)|S_s,\omega(a_s))&=&\int_0^S\,P(S',\omega(a')|S_s,\omega(a_s))\d S'
\nonumber\\
&=&\frac{1}{2}\frac{\omega(a_s)-\omega(a)}{\omega(a_s)}{\rm Exp}\left[\frac{2\omega(a)(\omega(a_s)-\omega(a))}{S_s}\right]{\rm Erfc}(X)+\frac{1}{2}{\rm Erfc}(Y),\nonumber\\
X&=&\frac{S(\omega(a_s)-2\omega(a))+S_s\omega(a)}{\left[2 S_s S(S_s-S)\right]^{1/2}}\nonumber\\
Y&=&\frac{S_s\omega(a)-S\omega(a_s)}{\left[2 S_s S(S_s-S)\right]^{1/2}}.
\end{eqnarray}
Assuming that $S\ll S_s$, $\omega(a)\ll \omega(a_s)$, but $(S_s/S) =(\omega(a_s)/\omega(a))^2$, it is easy to see that $X\simeq Y\simeq \omega(a)/\sqrt{2 S}$ the above probability is 
again suppressed, 
\begin{align}
P(S,\omega(a)|S_s,\omega(a_s))&= 
\frac{1}{2} {\rm Erfc} \lp \frac{\omega(a)}{\sqrt{2 S}} \rp 
\llp 1+ \exp \lp 2 \frac{ \omega (a)}{\omega (a_s)} \frac{\omega ^2(a_s)}{S_s}\rp \rrp
\nn \\
&\simeq \beta(M(r_{m}))\ll 1,
\end{align}
where we used the fact that $\omega(a_s)/\sqrt{S_s} = {\cal O}(6 \div 8)$, and the argument in the exponential enhancement factor is teamed by the small coefficient $\omega(a)/\omega(a_s) \ll 1$.

This confirms the intuitive picture that, since formation of a PBH is a rather rare event, having a small PBH swallowed by a larger one is even rarer and it goes like
$\beta(M(r_{m}))\beta(M(r_{m,s}))\simeq \beta^2(M(r_{m,s}))$, where in the last passage we have used the fact that for a broad spectrum the abundance is the same for small and big PBHs (since $\omega^2(a)/S=\omega^2(a_s)/S_s)$).

Having established the absence of the cloud-in-cloud problem for broad flat spectra, and since the abundance $\beta$ is the same for all masses, a natural question is
then what is the final mass distribution. To answer this question we recall that 
\be
R\sim k^{-1}\sim  1/aH\sim a\sim M_H^{1/2}.
\ee
Taking into account that the density of small-mass PBHs grows like non-relativistic matter, and that $\beta=\rho_\pbh/\rho_{\rm tot}$ at the time of the PBH formation, the abundance of the small PBHs at the formation time $t_l$ of the large ones is 
\bea
	\frac{\rho_s (t_l)}{\rho_l (t_l)} &=& \frac{\rho_s (t_s) (a_s/a_l)^3}{\rho_l (t_l)}
	=
	\frac{\rho_s (t_s) (a_s/a_l)^3}{\rho_l (t_l)} \frac{\rho_{\rm tot} (t_l) a_l^4}{\rho_{\rm tot} (t_s) a_s^4}
	\nn \\
	&=&\frac{\beta(M(r_{m,s}))}{\beta(M(r_{m,l}))} \frac{a_l}{a_s}
	=\frac{\beta(M(r_{m,s}))}{\beta(M(r_{m,l}))} \frac{k_s}{k_l}= \frac{k_s}{k_l}=\left(\frac{M(r_{m,l})}{M(r_{m,s})}\right)^{1/2}\gg 1.
\eea
Therefore, even though the probability of forming a PBH is independent of the mass, i.e. $\beta(M(r_{m,s}))=\beta(M(r_{m,l}))$, the PBHs with the smallest  mass will give the largest contribution to the mass distribution. In this sense,  again, the predictions are  similar to the ones of a narrow spectrum. 

Of course if the hierarchy between the smallest and largest momenta is not that large, the big PBHs could have a non-negligible abundance and  play an interesting cosmological role.

%
 
 \subsection{Broad spectra with a red tilt}
If the spectrum is broad, but with a red tilt, $n_p<0$, from the expression (\ref{conds}) one deduces that large-scale variances computed at their horizon crossing time $t_l$ are larger than those at small scales when they entered the horizon. The formation of PBHs will be therefore dominated by the largest scales, and PBHs will have the largest possible mass allowed by the power spectrum. In such a case, the presence of the small PBHs is highly suppressed, and the cloud-in-cloud problem is absent. In this sense, broad spectra with a red tilt also give predictions similar to the peaked power spectra.
 
%
One could argue that  if the broad spectrum is red or flat, the production of the lighter PBHs could be significantly enhanced at later times by the impact of the long modes perturbations on the small ones after they have entered the horizon \cite{ser}. However, long modes cross the horizon when the short modes are well inside the horizon. By that time, the radiation pressure has already dispersed the small scales peaks (see,  for example,  Ref. \cite{rez} for a discussion of the dynamics of the sub-threshold peaks upon horizon re-entry).
 This effect  is captured already at the linear level by the transfer function in Eq. (\ref{tf}) in a radiation-dominated universe, which scales like $1/(k\eta)^2\ll 1$ on subhorizon scales.
Therefore short-scale variances promptly decay in contrast to what was suggested in Ref. \cite{ser}. The time evolution of each mode prevents different scales from sizeably impacting each other.  
 
 Another option is that PBHs are also generated at small-scales. This will require though that $\sigma_l\gg\sigma_s$ and the danger is to enter a  non-linear regime before the long mode enter the horizon. We will come back to these considerations in the section devoted to the clustering.
 
 \section{The supreme statistics}
 \noindent
 In this section we return to the case of a broad spectrum as in Eq.~\eqref{eq:broad_specc} with a blue tilt $(n_p>0)$ and we wish to show with a different technique, the supreme statistics, that small PBHs are the only ones which are relevant. The reader should be aware of the fact that in this section we return to work with the physical quantities and not with  the rescaled ones.
 
Consider  a spherical peak (we consider for simplicity only spherical peaks, which is a good approximation for  those rare peaks giving rise to PBHs) whose typical scale is $r_{m,l}\gg r_{m,s}$. This peak re-enters the horizon when $a_l H_l r_{m,l}\sim 1$, and its volume-averaged density contrast is
\be
\delta_{l}=\frac{4\pi}{V_{l}}\int_0^{r_{m,l}}{\rm d}r\, r^2\, \delta(r,t),\quad V_{l}=\frac{4\pi}{3}r^3_{m,l}.
\ee
 Inside this large spherical peak of size $r_{m,l}$ there are about
\be
N\simeq (r_{m,l}/r_{m,s})^3
\ee 
subpeaks with volume of $4\pi r_{m,s}^3/3$ and typical scale of $r_{m,s}\sim 1/k_s$. The volume-averaged density contrast characterising each subpeak is
\be
\delta_{s}=\frac{4\pi}{V_{s}}\int_0^{r_{m,s}}{\rm d}r\, r^2\, \delta(r,t),\quad V_{s}=\frac{4\pi}{3}r^3_{m,s}.
\ee
These subpeaks re-enter the horizon when $a_s H_s r_{m,s}\sim 1$, with $ a_s H_s\gg  a_l H_l$, therefore they re-enter the horizon much earlier than the large peak does. The critical threshold for the formation of a PBHs is the same for the large peak and its subpeaks, while the variances are different at each horizon crossing $t_s$ and $t_l$ repsectively (we do not consider  here the radiation transfer function)
\bea
\label{condsss}
\sigma_s^2 (t_s)&\simeq& \frac{16}{81(n_p+4)}{\cal P}_0\left(\frac{k_s}{a_sH_s}\right)^4\simeq \frac{16}{81(n_p+4)}{\cal P}_0,\nonumber\\
\sigma_l^2 (t_l) &\simeq& \frac{16}{81(n_p+4)}{\cal P}_0\left(\frac{k_l}{a_lH_l}\right)^4\left(\frac{k_l}{k_s}\right)^{n_p}\ll \frac{16}{81(n_p+4)}{\cal P}_0.
\eea

The question we wish to answer is whether the subpeaks form PBHs more easily than the large peaks. To answer this question we can use  the so-called supreme statistics (or extreme-value statistics) \cite{bar} to study the statistics of the highest subpeaks, and determine the largest value of the subpeak density contrast at the short scale. It is this highest subpeak which will determine if a collapse into a PBH takes place on the short scale. In more general terms, considering $n$ samples, each of size $m$, all drawn from the same underlying distribution (which is Gaussian in our case), the distribution of the maxima within each sample, and therefore the most probable maximum value, can be determined using the supreme statistics \footnote{Of course in the limit $n\to\infty$ one recovers the Gaussin distribution and the largest value goes to infinity.}. 
\begin{figure}[t!]
	\centering
		\includegraphics[width=.35\textwidth]{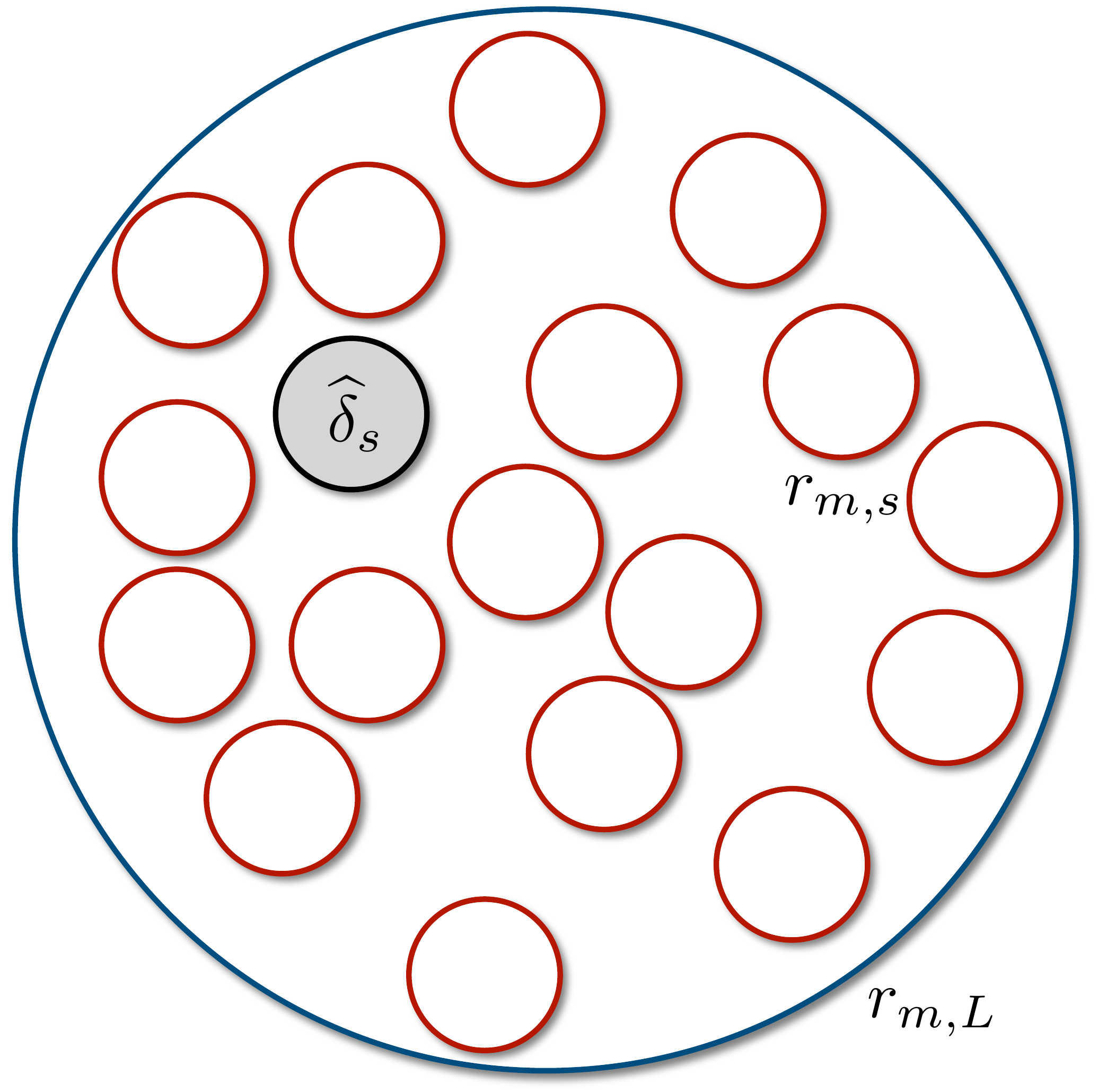}
		\caption{\it A schematic picture of the subpeaks within a large spherical peak.}
		\label{subpeaks}
		\end{figure}
\noindent

In our case we have $m=1$ and $n=N$. Specializing to a Gaussian parent distribution, the probability that the density in any one subpeak is less than a given value $\widehat{\delta}_{s}$ is given by the cumulative distribution function, 
\begin{eqnarray}
P_1(\widehat{\delta}_{s}) &=& \int _{-\infty}^{\widehat{\delta}_{s}} P(\delta_{s}|\delta_{l}) \d \delta_{s}\nonumber\\
&=&1-\frac{1}{2}{\rm Erfc}\left(\frac{\widehat{\delta}_{s}-\delta_{l}\frac{\langle\delta_{l}\delta_{s}\rangle}{\sigma^2_{l}}
}{\sqrt{2}\sqrt{\sigma^2_{s}-\frac{\langle\delta_{l}\delta_{s}\rangle^2}{\sigma^2_{l}}}}\right),
\label{p1}
\end{eqnarray}
where $P(\delta_{s}|\delta_{l}) $ is the conditional probability for the density at short-scales $\delta_s$, given the density at large scales $\delta_l$. Since both $\delta_s$ and $\delta_l$ are Gaussian distributed, their conditional probability is also a Gaussian. For large values of $\delta_s$ and $\sigma_s$, this expression can be safely approximated to 
\be
P_1\simeq 1-\frac{1}{2}{\rm Erfc}\left(\frac{\widehat{\delta}_{s}}{\sqrt{2}\sigma_{s}}\right),
\ee
and hence
\be\label{pdf_ld}
P(\widehat{\delta}_{s}|{\delta}_{l}) = \frac{dP_1}{d\delta_s} \approx \frac{1}{\sqrt{2\pi \sigma^2_{s}}} \exp \left\{ -\widehat{\delta}^2_{s}/2\sigma^2_{s}\right\},
\ee
The probability that the density of all $N$ subpeaks is less than $\widehat{\delta}_{s}$ is given by 
\begin{equation}
\label{eq:peak_prof}
P_N(\widehat{\delta}_{s}) = P_1^N(\widehat{\delta}_{s}),
\end{equation} 
and the differential probability that a subpeak of density $\widehat{\delta}_{s}$ at a scale $r_{m,s}$ has the highest density among all the subpeaks is given by $\d P_N/\d  \widehat{\delta}_{s}$. From this probability distribution, we can compute the mean (and variance) of the full set of subpeaks. In figure \ref{subpeaks}, we show a schematic picture of the subpeaks within a large spherical peak.

In the limit of large $\widehat\delta_s$, one can obtain a simple analytic expression for the most probable value of $\widehat\delta_s$, as was done in Ref. \cite{us}.  The maximum of $\d P_N/\d  \widehat{\delta}_{s}$ determines the most likely value of $\widehat{\delta}_{s}$,
\be
\frac{\d^2P_1^N}{\d\widehat{\delta}^2_{{s}}}  = (N-1) \left(\frac{\d P_1}{\d \widehat{\delta}_{s}} \right)^2 + P_1 \frac{\d^2 P_1}{\d \widehat{\delta}^2_{s}}=0.
\label{detbar}
\ee
Since by definition $\d P_1/\d  \widehat{\delta}_{s}=P(\widehat{\delta}_{s}|\delta_{l})$, and $P_1$ is nearly one in the large-$\widehat \delta_s$ limit, Eq. \eqref{detbar} reduces to
\be
(N-1) P^2(\widehat{\delta}_{s}|{\delta}_{l}) = -\frac{\d P(\widehat{\delta}_{s}|{\delta}_{l})}{\d \widehat{\delta}_{s}},
\label{start}
\ee
which using Eq. \eqref{pdf_ld} gives 
\be
\frac{N-1}{\sqrt{2\pi}} e^{-\widehat{\delta}^2_{s}/2\sigma^2_{s}} = \frac{\widehat{\delta}_{s}}{\sigma_{s}}.
\ee
Therefore, the exponential suppression is offset by the large number of subregions that can attain independent values. Since $N\simeq (r_{m,l}/r_{m,s})^3\gg 1$, to a good approximation we finally find \cite{us}
\be\label{avde}
\widehat{\delta}_{s}\simeq \sqrt{2} \,\sigma_{s}\,\ln^{1/2} N=\sqrt{6} \,\sigma_{s}\,\ln^{1/2} (r_{m,l}/r_{m,s}),
\ee
which, as expected, for $N\to\infty$ goes to infinity.  Note that the density on small-scales does not depend on the boundary conditions, i.e. the large-scale density. Physically, this can be understood as that the very small scales fluctuations are so large that they are virtually independent of the large-scale density field.

\begin{figure}[t!]
	\centering
	\includegraphics[width=.495\textwidth]{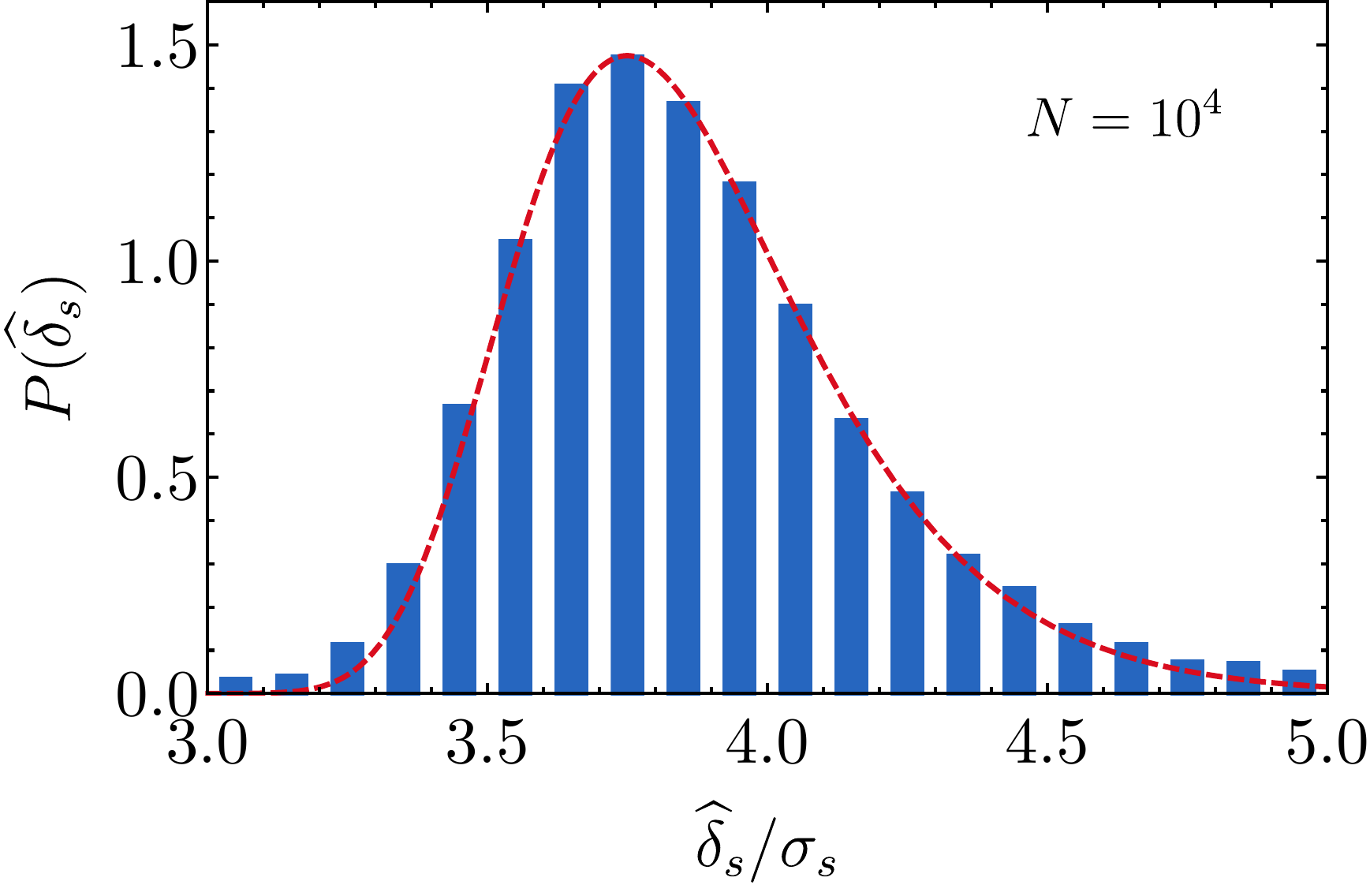}
		\caption{\it Comparison of Eq.~\eqref{dalalpdf} with a numerical simulation with $ N=10^4$.
		}
		\label{fig2}
\end{figure}

In fact, a more refined computation for the limit of large $N$, shows that the probability distribution of the largest value $\widehat{\delta}_{s}$ is given by \cite{book,bar,dalal} 
\be\label{dalalpdf}
P(\widehat{\delta}_{s})\d \widehat{\delta}_{s}=\frac{\d}{\d \widehat{\delta}_{s}}
\left\{\exp\left[-\frac{N}{\sqrt{2\pi}}\frac{\sigma_{s}}{\widehat{\delta}_{s}}
e^{-\widehat{\delta}_{s}^2/2\sigma^2_{s}}
\right]\right\}\d \widehat{\delta}_{s} .
\ee
A comparison with a numerical simulation of several  samples of $N=10^4$ is shown in figure \ref{fig2}.
\noindent
The key point is  that, even in the presence of a broad peak with size $r_{m,l}$, there is on average one subpeak at a given scale $r_{m,s}$ whose value of the overdensity is 
given by the maximal value $\widehat{\delta}_{s}\gg \delta_{l}$.  It is therefore at the smallest scale $r_{m,s}$ that the PBH forms first once the comoving horizon grows up to a value $\sim r_{m,s}$. In other words, at the re-entry of the small scale modes into the comoving Hubble horizon, if the corresponding smoothed overdensity is large enough, the PBHs will form.

One would conclude that, if  $\widehat{\delta}_{s}$ is above (and close to) the threshold, one expects to form roughly one PBH in each volume $V_s$ (with $\widehat{\delta}_{s}$ being the maximum which is realized only in one out of the $N$ samples), that is 
\begin{equation}
\frac{ 	{\cal O}(1)\,  \text{PBH}} {V_l}=
\frac{	  {\cal O}(1)\,\text{PBH}} { N V_s}.
\end{equation}
The mass fraction composed of the PBHs with mass $M(r_{m,s})$ is therefore $(x=\widehat{\delta}_{s}/\sigma_{s}$)
\begin{align}\label{mf}
	\beta &= \frac{f_{l}}{N} \int_{\delta_c/\sigma_{s}}^{\infty}  \frac{\d }{\d x} \left \{  \exp \llp - \frac{N}{\sqrt{2 \pi }} x^{-1} e^{- \frac{x^2}{2}} \rrp  \right\} \d x
	\nonumber \\
	&=
	\frac{f_{l}}{N}
	\llp \exp \llp - \frac{N}{\sqrt{2 \pi }} x^{-1} e^{- \frac{x^2}{2}} \rrp_{x\to \infty } -
	\exp \llp - \frac{N}{\sqrt{2 \pi }} x^{-1} e^{- \frac{x^2}{2}} \rrp _{x\to \delta_c/\sigma_{s} }\rrp
\nonumber	\\
& =\frac{f_{l}}{N} \llp 1- \lp 1 - \frac{N \sigma_{s}}{\delta_{c} \sqrt{2\pi}} e^{-\delta^2_{c }/2 \sigma^2_{s}} +
\dots
 \rp  \rrp
\nonumber	\\
& =\frac{f_{l}}{N} \llp  \frac{N  \sigma_{s}}{\delta_{c} \sqrt{2\pi}} e^{-\delta^2_{c }/2 \sigma^2_{s}} +\dots  \rrp\nonumber\\
&\simeq f_{l}\beta(M(r_{m,s})),
\end{align}
regardless of the value of $N$. Here the  factor $f_{l}$ accounts for the number density of the larger peaks. 

The result in Eq. \eqref{mf} summarizes our point. Even if the initial power spectrum of the curvature perturbation is broad,  if it grows with the momentum, the population of PBHs is  dominated by those which form first, i.e. from the smallest scale fluctuations. Furthermore, the cloud-in-cloud probability for these small PBHs to be swallowed by larger PBHs is very tiny. In  our final expression, this results in the further suppression factor $f_{l}$, which is extremely small, given that the typical value of $\delta_{l}$ is much smaller than $\delta_c$. Note that summing over all peaks of size $r_{m,l}$, even those not giving rise to PBHs, the factor $f_{l}$ tends to unity and one recovers the standard  abundance of PBHs with mass $M(r_{m,s})$. 

To summarize, our expectation is that when  larger modes re-enter the horizon, the PBHs of higher masses do not start forming, neither do they engulf the previously formed lighter PBHs. This result reenforces the conclusion in the previous section, which was based on the excursion set method applied to PBHs.

\section{The clustering at formation}
\noindent
Another relevant issue arising when PBHs are generated by broad spectra is whether or not they are clustered at the formation time. Initial clustering can affect  PBH merging and accretion, leading to different spatial distributions at later times. This in turn can weaken (or strengthen) the current observational constraints on their  abundances.  

Being discrete objects, the   two-point correlation function of PBHs have the   general form
\begin{eqnarray}
\big\langle\dpbh(\vr)\dpbh(\vec{0}) \big\rangle 
=
\frac{1}{\npbh}\delta_D(\vr)+ \xipbh(r),
\label{eq:PBH2pt}
\end{eqnarray}
where  $\delta_D(\vr)$ is the three-dimensional Dirac distribution, $\npbh$ is the average number density of PBH per comoving volume,   and $\xipbh(r)$ is the reduced PBH correlation function. In Ref. \cite{v} it was shown that for narrow spectra with amplitude ${\cal P}_0$, in the allowed range of PBH masses, the comoving scale $\xxi$ at which $ \xipbh(r)\simeq 1$ is always smaller than the average comoving separation between two PBHs,  $\overline{r}=(3/4\pi\npbh)^{1/3}$ 
  \begin{eqnarray}
  \frac{\xxi}{\overline{r}} &\simeq& 10^{-3}\left(\frac{{\cal P}_0}{0.1}\right)^{1/4}\left(\frac{\mpbh}{\msun}\right)^{5/6},\nonumber\\
 \xxi &\simeq& 10^{-7} \left(\frac{{\cal P}_0}{0.1}\right)^{1/4}\left(\frac{\mpbh}{\msun}\right)^{1/2}\,{\rm Mpc}.
  \end{eqnarray}
For instance, for the  case of PBHs with $\mpbh\sim 30 \msun$, the initial distances relevant for the present merger rate is $\gsim 4\cdot 10^{-5}$ Mpc \cite{ser}, and  it was concluded that for narrow spectra clustering is not relevant.  

In the previous sections we have argued that, since PBHs are rare events, the cloud-in-cloud phenomenon is basically absent. Furthermore,  the fact that the final spectrum of PBHs is tilted either towards small-scales PBHs (for flat broad spectra or blue-tilted broad spectra) or large-scale PBHs (for red-tilted broad spectra) the expectation as far as the mass function is concerned is similar to that of a narrow power spectrum. 

These two considerations  suggest that the clustering properties at formation will not differ from the case of narrow spectra. In other words, the PBHs will not be clustered. As already mentioned, we do not expect this conclusion to be altered by the possibility, in the case of flat or red-tilted broad spectra, that the formation of small-scale PBHs is  enhanced at later times by the effect of the long modes perturbations once the latter have  entered the horizon.  Indeed, by that time the small-scale peaks with an amplitude below the threshold (at the time when they enter the horizon) have been already washed-out by the radiation pressure. 

We use the supreme statistics again to elaborate further on the clustering for the case of a broad spectrum with a blue tilt. Having argued that for broad spectra with a blue tilt PBHs form at the smallest scale without being swallowed by  the more massive PBHs, one can ask if nevertheless the correlation length of the lighter PBHs is larger than the Poisson distance. At the time of formation the average comoving separation between such PBHs is
\be
\overline{r}_s\simeq \frac{1}{\beta^{1/3}(M(r_{m,s}))a_s H_s},
\ee
provided that  only one subpeak within the spherical peak of radius $r_{m,L}$ collapses to form a PBH (this is the assumption made in the previous section). However, if more than one subpeak collapses into a PBH the correlation length will be given roughly by $r_{m,l}$. For instance, one can compute the probability distribution function of the
$p$-th maximum out of a sample of size $N$, which is given by \cite{book2}
\begin{equation}
 P_p(x) \d x = \frac{N!}{(N-p)!(p-1)!} \llp\int_{-\infty}^{x} f(x') \d x'\rrp ^{N-p}
 \llp\int_{x}^{\infty} f(x') \d x'\rrp ^{p-1}
 f(x) \d x 
\end{equation}
where $f(x) = \exp(-x^2/2)/\sqrt{2\pi}$ is the Gaussian probability distribution function. It is easy to show that, in the case of the first maximum $p=1$, the previous distribution recovers the one in Eq.~\eqref{dalalpdf} in the large $N$ limit. 

\begin{figure}[t!]
	\centering
	\includegraphics[width=.445\textwidth]{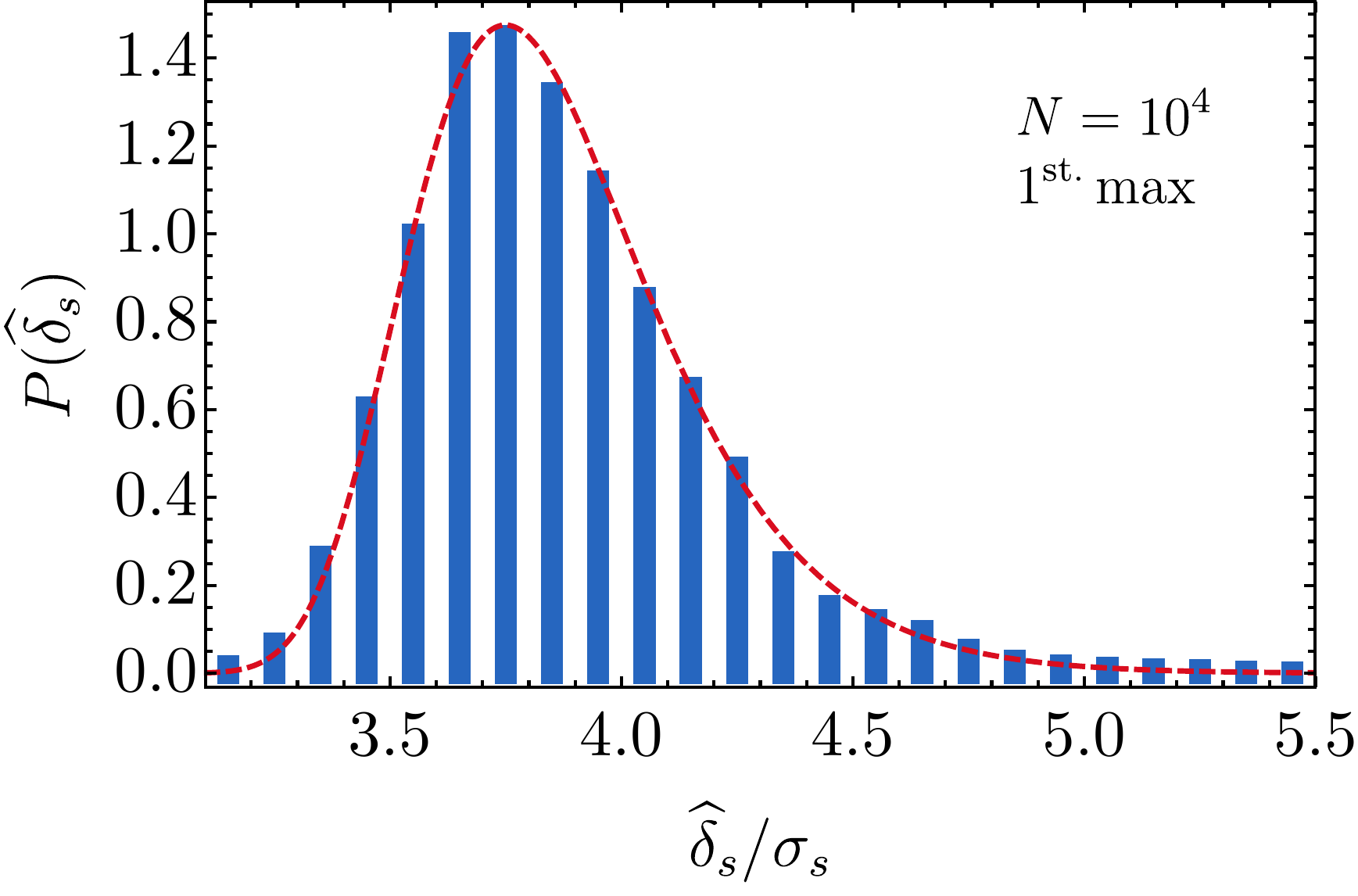}
	\includegraphics[width=.44\textwidth]{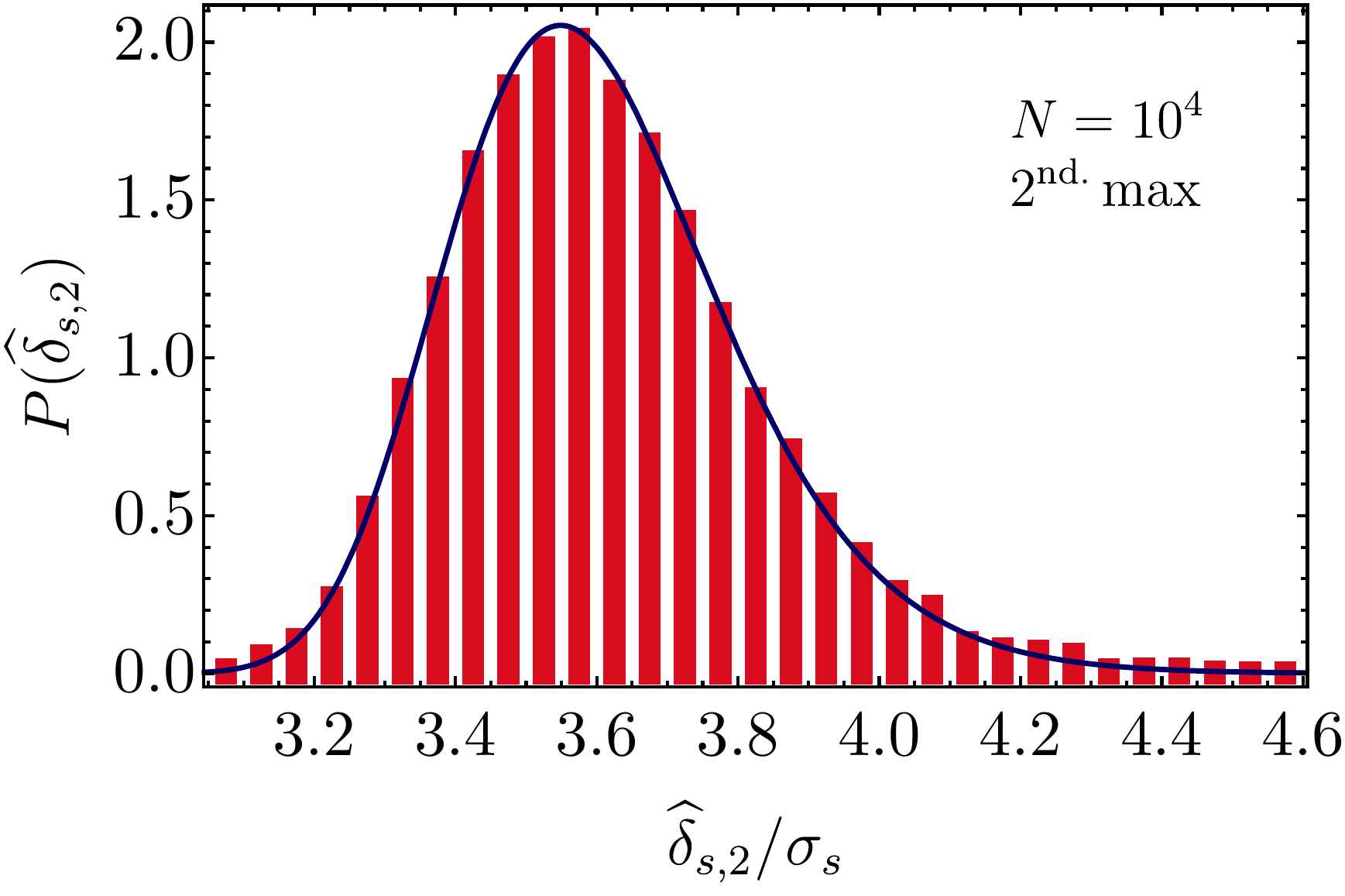}
		\caption{\it \textbf{Left:} Probability distribution function for the first maximum out of $N$ element Gaussian distributed as in  Eq.~\eqref{dalalpdf} and its comparison with a numerical simulation of $N$ realisation of samples with $N$ elements each. \textbf{Right:} probability distribution function of the second maximum and its comparison with a numerical simulation.} 
		\label{fig9}
\end{figure}

In order to deal with more tractable analytical formulae one can also  approximate the probability $ P_p(x)$ with the following form  (we have checked that for large $N$ the results coincide) \cite{bar}
\begin{equation}\label{bar-pdf}
	P_p(x) \d x =  \frac{ p^p a_p }{(p-1)!}\exp \llp  - p y_p -pe^{  -y_p } \rrp  \d x
	\qquad \text{with}\qquad
	y_p = a_p (x-x_p),\quad x=\widehat{\delta}_{s,p}/\sigma_s,	
\end{equation}
where $x_p$ is the mode of the probability, where the latter has its maximum. For $p=1$ one recovers  Eq.~\eqref{dalalpdf}. As for the coefficients, we have found them numerically, see Fig.~\ref{fig9},  and find, for instance, that  $a_1\simeq a_2 \simeq  \log_{10} N$.
In the limit of large $N$ and small $p$, the distance between different ranked maxima, defined as
\begin{equation}
	d_{p}= x_{p+1} -x_{p}, \qquad \text{for} \qquad 1\geq m \geq n-1, 
\end{equation}
is distributed as
\begin{equation}
	f(d_p) = p \,a_{p+1} \exp \llp -p \,d_p \,a_{p+1}\rrp. 
\end{equation}
Therefore, knowing the expectation value for the first ranked maximum, we can infer the expectation value for the second and so on. In particular, the most probable $d_m$ is
\begin{equation}
 \langle d_p \rangle  = \int_0 ^\infty  \d d_p\,d_p \,f(d_p) = \frac{1}{p \,a_p }. 
\end{equation}
For the second maximum we therefore get 
\begin{equation}\label{expsec}
	\widehat{\delta}_{s}-\widehat{\delta}_{s,2}\sim \frac{\sigma_s}{a_1}\simeq \frac{\sigma_s}{\log_{10} N}.
\end{equation}

\begin{figure}[t!]
	\centering
	\includegraphics[width=.49\textwidth]{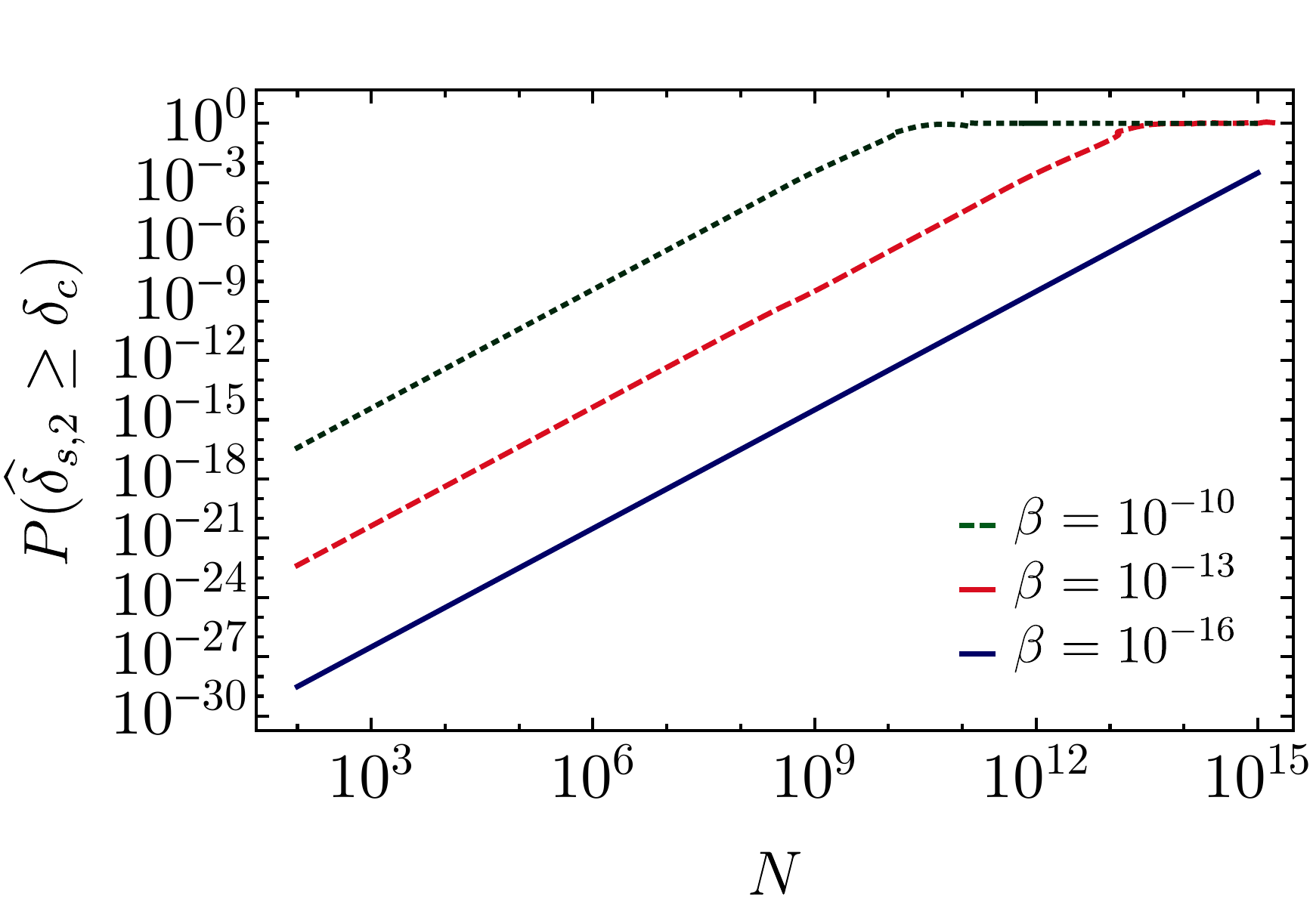}
		\caption{\it Probability of having $\widehat \delta_{s,2} \geq \delta_c$ depending on the value of $N$ considering three different $\beta$.} 
		\label{cl}
\end{figure}

Using the above distribution we can compute the probability to have two PBHs in a peak of size $r_{m,l}$. This is equivalent to computing the probability that the second maximum is above or equal the threshold $\delta_c$ as
\begin{equation}
P(\widehat \delta_{s,2} \geq \delta_c) =	\int_{\delta_c/\sigma_s}^\infty \d x\,P_2(x).
\end{equation}
 The result is shown  in Fig.~\ref{cl}. We stress the dependence of $P(\widehat \delta_{s,2} \geq \delta_c)$ on the variance $\sigma_s$ through the lower integration limit.
\noindent

Suppose now that the broad spectrum extends over a range of scales such that $r_{m,l}\simeq N^{1/3}r_{m,s}\simeq N^{1/3}/a_sH_s$. 
In order for the distance of these two subpeaks to be smaller than the Poissonian one, the following condition must be imposed
\be
\frac{N^{1/3}}{a_sH_s}\lsim  \frac{1}{\beta^{1/3}(M(r_{m,s}))a_s H_s}\quad{\rm or}\quad N\lsim \frac{1}{\beta}\simeq 10^{8}\left(\frac{M_\odot}{M}\right)^{1/2},
\ee
where in the last passage we have assumed that the PBHs form the dark matter. On the other hand from  Fig.~\ref{cl} we learn that the probability of having two subpeaks above the threshold is negligibly small unless $N\gsim \beta^{-1}$.  This can be understood by realizing that the two maxima are very close to each other and therefore
 \begin{equation}
P(\widehat \delta_{s,2} \geq \delta_c) \simeq \int_{\delta_c/\sigma_s}^\infty \d x\,P_1(x)\simeq N\,\beta,
\end{equation}
This indicates that the correlation length is never smaller than the average Poisson distance for a broad spectrum with a blue tilt,  suggesting that  PBHs are not clustered at formation.
%
%


\section{Conclusions}
\setcounter{section}{5}
\noindent
In this paper we have examined the formation of PBHs that are generated by sizeable small-scale curvature perturbations originated during inflation, upon horizon re-entry. We have particularly focused on the case of broad spectra for the curvature perturbation in which one may not immediately identify a single mass-scale to associate with the PBHs. By using the excursion set method we have argued that the probability of the small-mass PBHs being absorbed by the bigger ones is negligible. This is due to 
the fact that the formation of PBHs is by itself a rare event. This result has been confirmed by using the supreme statistics which describes the distribution of extreme values.

Our findings indicate that the predictions from broad spectra are qualitatively the same as for narrow spectra. The mass distribution of the PBHs is dominated by a single mass (which can be associated to the maximum or the minimum of the momenta of the power spectrum depending on its tilt). In particular, we have offered indications that the PBHs at formation 
are not clustered. Of course the subsequent evolution of the clustering will depend upon the merging and accretion of the PBHs, and might be fully  captured only by real-time numerical simulations. Another aspect worth investigating more in detail is the role of non-Gaussianities in our considerations. For instance, the presence of local-type primordial non-Gaussianity might enhance the clustering on large scales \cite{ng1,Suyama:2019cst} due to the correlation of small and large scales. 

\begin{center}
{\bf  Acknowledgments}
\end{center}
\noindent
We thank V. Dejaques, I. Musco and S. Young for many useful discussions. The authors  are  supported by the Swiss National Science Foundation (SNSF), project {\sl The Non-Gaussian Universe and Cosmological Symmetries}, project number: 200020-178787.



\end{document}